# An intermediate baryon system formation and the angular distributions of the slow particles emitted in hadron-nuclear and nuclear-nuclear interactions at high energies

M K Suleymanov[1,2,*], O B Abdinov[3], R M Aliyev[3], F M Aliyev[3], M Q Haseeb,
Y H Huseynaliyev[2,5], E U Khan[2], A Kravchakova[4], E I Shahaliev[1],
S Vokal[1,4] and A S Vodopianov[1]

[1] Veksler and Baldin Laboratory of High-Energies, Joint Institute for Nuclear Researches, Dubna, Russia
[2] COMSATS Institute of Information Technology, Islamabad, Pakistan
[3] Institute of Physics, Academy of Sciences, Baku, Azerbaijan
[4] University of P. J. Shafarik, Koshice, Slovak Republic
[5] Department of Physics, Sumgayit State University, Azerbaijan

* E-mail: mais@jinr.ru

**Abstract.** We have analyzed the angular distributions of the b- particles emitted in Kr+Em -reaction at 0.95 A GeV and in Au+Em -reaction at 10.7 A GeV and compared these with lighter projectile experiments for which some structure in the angular distribution of slow particles was observed. The same structure for the b-particles almost disappears by increasing the projectile mass. We believe that it is connected with increasing rates of internuclear secondary interactions which could lead mainly to disappearance of the information about intermediate baryon system formation. We suggest that it should be taken into account at event generation for heavy ion interactions to restore the information on the intermediate baryon system formation.

## 1. Introduction

The experimental signals that give clue to the formation and decay of an intermediate baryon system [1-3] may be considered as important information to identify the new phases of nuclear matter. To consider the possibility of the appearance of deconfinement in cluster, Satz [4] discusses in detail that deconfinement is expected when the density of quarks and gluons becomes so high that it is no longer possible to separate colour-neutral hadrons, since these would strongly overlap. Instead clusters much larger than hadrons are formed, within which colour is not confined; deconfinement is thus related to cluster formation. This is the central topic of percolation theory, and hence a connection between percolation and deconfinement seems very likely [5].

Observation of the effects connected with formation and decay of the percolation clusters in heavy ion collisions at ultrarelativistic energies could be the first step for getting the information about deconfinement. Such signals may be obtained by studying the behaviour of the characteristics of secondary particles as a function of the centrality of collisions, mass and energy of the projectiles and targets. The formation and decay of the intermediate baryons could influence the characteristics of secondary particles, especially those of slow particles. Ref. [6] discusses that the angular distributions of fragments could have special structure as a result of the formation and decay of some intermediate states.

What we have got from experiments? There are only a few experimental results [7-10] that could be considered as a possible confirmation of it. Let us consider them somewhat in detail. Anoshin et al [7] have observed that the angular distributions of protons emitted in $\pi^{-12}$C-interaction (at 40 GeV/c) in central collisions (with total disintegration of nuclei) have some structure with peak at angles close to $60^0$. This result was confirmed in angular distributions of protons emitted in $\pi^{-12}$C -interaction with total disintegration of nuclei at 5 GeV/c [8]. In Ref. [7, 8] protons with momentum less than 1 GeV/c were considered. Hecman et al [9] study the angular distributions for the slow protons emitted in the central He+Em (emulsion) - (at 2.1 A GeV), O+Em - (2.1 A GeV) and Ar+Em - (1.8 A GeV) collisions. Some wide structure was observed in these distributions. Almost similar but wider structures have been observed in the angular distributions of the black or b-particles (mostly protons with $p \leq 0.2$ GeV/c and multiple charged target fragments having a range, 3mm) emitted in the Ne+Em -reactions at 4.1 A GeV [10] and reported that the structure becomes cleaner in central collisions.

The main result of these experiments is that the angular distributions of slow particles emitted in π-meson, proton and light nuclei interactions with nuclear targets indicate some structure. The reason of this structure could be the formation and decay of an intermediate baryon system, e.g. percolation clusters (see [11]).

Now what we have got for angular distributions of slow particles emitted in intermediate and heavy ion collisions? Up to now those distributions could not show any structure. Why? May be it is because the intermediate baryon systems have a small cross section (statistical reason) or the information on these formations is lost for reasons, such as the secondary multiparticle interaction, scattering and rescatterings (dynamical reason). We can exclude the statistical reason because the structure for slow particle's angular distributions has been observed with lesser statistical data. What about dynamical reason?

To get an answer to this question we study in detail the angular distribution of slow particles emitted at intermediate mass and heavy nuclear interaction with emulsion target at relativistic energies.

### 2. Angular distribution of the b-particles emitted in intermediate mass and heavy nuclei interaction

To analyze angular distribution of the b -particles as a function of centrality [12] the experimental data on Kr+Em -reaction at 0.95 A GeV [13] and Au+Em - reaction at 10.7 A GeV [14] have been used. The experimental data using nuclear beams of the SIS GSI and AGS BNL by the EMU01 Collaboration were obtained. All 842 Kr+Em events and 1185 Au+Em ones have been considered.

The angular distributions of slow particles were presented by the EMU01 Collaboration (for example see [15]). However the angular distributions of the slow fragments are considered by us separately for:

1) all events and the events with a number of $N_h \geq 8$ ($N_h$ is a number of h-particles). This criterion has been used to separate the heavy nuclei interaction. The h -particles are sum of g (grey particles are those whose ionization, the number of grains per unit length, correspond to protons with momentum $0.2 \leq p \leq 1.0$ GeV/c) and b -particles;

2) peripheral and central collisions.

The experimental data has been compared with data generated by the cascade evaporation model (CEM) [16].

Fig 1 a-d shows the angular distributions of the b- particles emitted in the Kr + Em - reactions at 0.95 A GeV. Herein and further all distributions were normalized on a total square under the curves. The results are shown separately for all events (Fig.1a) and the events with $N_h \geq 8$ (Fig.1b). The results from the CEM have also been shown in these figures. We cannot see any structure in these distributions that were discussed in [7-10].

In Fig 2 a-d the angular distributions of the b- particles emitted in the Au + Em - reactions at 10.7 A GeV are shown. As in Fig. 1 a-b, the results are demonstrated separately for all events (Fig.2a) and the events with $N_h \geq 8$ (Fig.2b). These figures show some structure in the angular distributions for b-particles that is cleaner for the all events for which the rates of the peripheral collisions are stronger. The possible reasons for the appearance of this structure are the multinucleons interactions, scattering and rescatterings effects.

In Fig 3 a-b the angular distributions of the b- particles emitted in the Kr + Em - reactions at 0.95 A GeV are demonstrated. The results for the central collisions (Fig 3a) and the peripheral ones (Fig 3b) are shown separately. To select the central collisions, we used the criteria $N_g \geq 20$ suggested by Abdinov et al [17]. One can see that the angular distributions are completely described by CEM.

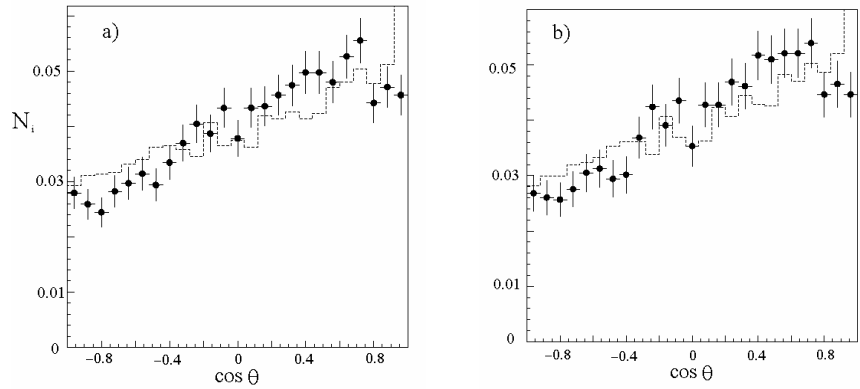

**Figure 1 a-b** Angular distributions of the b- particles emitted in Kr+Em-reactions at 0.95 A GeV energy for all events - a); for the events with $N_h \geq 8$ – b). The histograms are the results coming from the CEM.

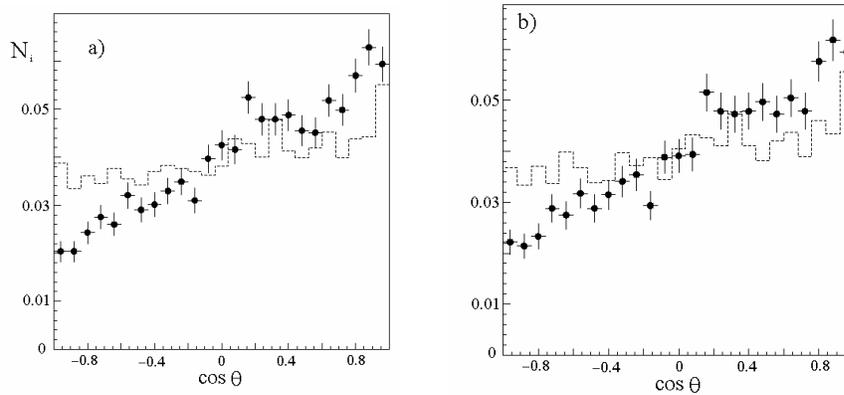

**Figure 2 a-b** Angular distributions of the b-particles emitted in Au+Em-reactions at 10.7 A GeV energy for all events - a); for the events with $N_h \geq 8$ – b). The histograms are the results coming from the CEM.

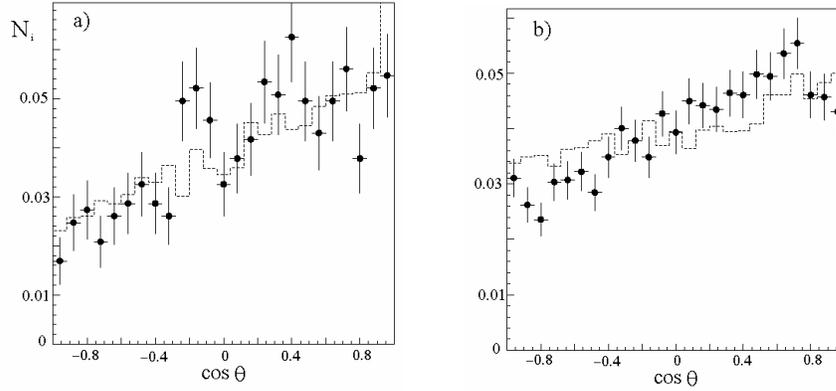

**Figure 3 a-b** Angular distributions of the b- particles emitted in Kr+Em-reactions at 0.95 A GeV energy for the central collisions - a); for the peripheral collisions – b). The histograms are the results coming from the CEM.

In the Fig 4 a-b the angular distributions of the b-particles emitted in the Au + Em - reactions at 10.7 A GeV are demonstrated. The results for the central collisions (Fig 4a) and the peripheral ones (Fig 4b) are shown separately. To select the central collisions, we used the criteria $N_F \geq 20$ though Abdinov et al [17] showed that the central Au+Em-events (at 10.7 A GeV) should be selected using the criteria $N_F \geq 40$. We followed a different criterion as the number of events is very small with $N_F \geq 40$. It is seen that in peripheral collisions the structure becomes cleaner (the structure diminishes and almost disappears in central collisions). Therefore as in Fig 2a-b one can say that the reason of the appearance of this structure is the multinucleons interactions, scattering and rescattering effects.

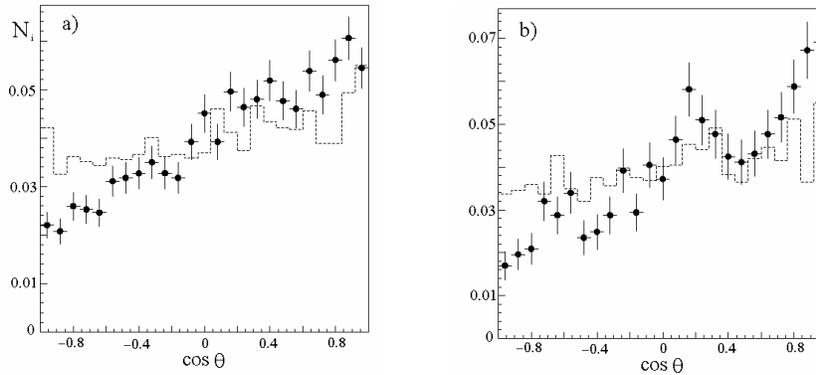

**Figure 4 a-b** Angular distributions of the b-particles emitted in Au+Em-reactions at 10.7 A GeV energy for the central collisions - a); for the peripheral collisions – b). The histograms are the results coming from the CEM.

### 3. Results and discussion

The angular distributions of b- particles emitted in Au+Em-reactions as well as in the Kr+Em-reactions do not contain any special structure except the one that is described using the usual mechanisms of the interaction (e.g. cascade evaporation). If we compare the results of the reactions mentioned above with those discussed in [7-10] one can see that by increasing the mass of the projectiles the angular distributions of slow particles change and the structure which was demonstrated in the case of π-mesons, protons and light nuclear projectiles almost disappears. This could be explained easily as follows: during the interaction of the heavier projectile with nuclear target, the number of secondary interactions as well as number of nucleon-nucleon elastic scattering and re-scattering events increases. These effects could lead to the disappearance of the information of any intermediate formations as well as the clusters.

But how we could restore this information? As already discussed, observation of the effects connected with formation and decay of the percolation clusters in heavy ion collisions at ultrarelativistic energies could be the first step for getting the information about deconfinement. We expected that such signals may be obtained by studying the behaviour of the characteristics of secondary particles as a function of the centrality of collisions, mass and energy of the projectiles and targets. But these could not be observed.

Therefore it is necessary to look for the ways for restoring the signal. One way could be using the generators of heavy ion events taking into account the results on the centrality dependences of other characteristics of secondary particles e.g. the average multiplicity of b-particles [17] which could be more sensitive to baryon system formation and decay. In Ref. [17] it was concluded that the formation of the percolation cluster could sufficiently influence on the behaviour of the slow particles average multiplicity as a function of centrality. Another way could be using the Fourier transformation and Maximum Entropy technique [18] to separate the spectrum of the signal. It is the subject of our investigations that would follow.

## Acknowledgements

A K and S V are indeed indebted for the support extended to us by the Agency for Science of the Ministry for Education of the Slovak Republic (Grant VEGA 1/2007/05). M K and Y H are thankful to the Higher Education Commission of the Islamic Republic of Pakistan (Grant N1-28/HEC/HDR/2006 and Grant N1-28/HEC/HRD/2007/501), CIIT (Islamabad) and JINR (Dubna).